\documentclass[aps,prd,onecolumn,groupedaddress,showpacs,nofootinbib,amssymb]{revtex4}
\usepackage{graphicx,bm,color}
\usepackage{amsmath}
\usepackage{amssymb}
\usepackage{amsfonts}
\usepackage{wrapfig}
\usepackage{cancel}

\newcommand{\be}{\begin{equation}}
\newcommand{\ee}{\end{equation}}
\newcommand{\bea}{\begin{eqnarray}}
\newcommand{\eea}{\end{eqnarray}}
\newcommand{\beaa}{\begin{eqnarray*}}
\newcommand{\eeaa}{\end{eqnarray*}}

\newcommand{\nn}{\nonumber \\}

\allowdisplaybreaks[4]

\begin{document}

\title{Sequestering Mechanism in Scalar-Tensor Gravity}

\author{Takuma Tsukamoto$^{1}$\footnote{
E-mail address: tsukamoto.takuma@h.mbox.nagoya-u.ac.jp}, 
Taishi Katsuragawa$^{2,3}$\footnote{
E-mail address: taishi@th.phys.nagoya-u.ac.jp},
Shin'ichi Nojiri$^{1, 2, 4}$\footnote{E-mail address:
nojiri@phys.nagoya-u.ac.jp}
}

\affiliation{
$^1$ Department of Physics, Nagoya University, Nagoya
464-8602, Japan \\
$^2$ Kobayashi-Maskawa Institute for the Origin of Particles and
the Universe, Nagoya University, Nagoya 464-8602, Japan \\
$^3$ Institute of Astrophysics, Central China Normal University, Wuhan 430079, China \\
$^4$ KEK Theory Center, High Energy Accelerator Research Organization (KEK), 
Oho 1-1, Tsukuba, Ibaraki 305-0801, Japan}

\begin{abstract}
We make a theoretical prediction for the ratio of the dark energy to other components in the Universe
based on the scenario of the sequestering mechanism \cite{Kaloper:2013zca,Kaloper:2014dqa,Kaloper:2015jra} 
which was recently proposed as one possible way to solve the cosmological constant problem.
In order to evaluate the value of dark energy and the others, 
we assume a specific scale factor which describes the big-crunch scenario in the scalar-tensor theory.
We specify the parameter region where the one can explain the observed dark energy-matter ratio. 
\end{abstract}

%\pacs{}

\maketitle

\section{Introduction \label{SecI}}

The quantum field theory is one of the pillars in the modern theoretical physics as it has provided the profusion of rigorous predictions for us in the various areas of the physics.
It is widely known that a strong divergence shows up in the quantum field theory
if one considers the quantum corrections from the matters to the vacuum energy. 
For example, the quantum correction from one bosonic degree of freedom
with mass $m$ is given by
\be
\label{01}
\rho_\mathrm{vacuum} 
= \frac{1}{\left( 2 \pi \right)^3}\int d^3 k \frac{1}{2} \sqrt{ k^2 + m^2 } \, .
\ee
The right-hand side of the above quantity diverges, and in order to regulate the divergence,
one may introduce the cutoff scale $\Lambda_\mathrm{cutoff}$,
to find that 
\be
\label{01B}
\rho_\mathrm{vacuum} \sim \Lambda_\mathrm{cutoff}^4 \, .
\ee
On the other hand, the above estimation of the vacuum energy gives us astounding result 
in the cosmology, 
where we have to face a notorious problem, so-called the cosmological constant problem.

The cosmological observations suggest that the cosmological constant $\Lambda$, 
which could be regarded as the vacuum energy in the Universe, is approximately 
equal to $\left( 10^{-3}\, \mathrm{eV} \right)^4$.
We find that the observed value of cosmological constant is much smaller than the value in (\ref{01B}) 
when we choose the cutoff scale $\Lambda_\mathrm{cutoff}$ 
to be the Planck mass scale $M_\mathrm{Planck}$;
assuming $\Lambda_\mathrm{cutoff} \sim M_\mathrm{Planck}$, we obtain 
\be
\label{I1}
\Lambda^{1/4} \sim 10^{-3}\,\mathrm{eV} \ll M_\mathrm{Planck} 
\sim 10^{19}\,\mathrm{GeV}=10^{28}\,\mathrm{eV}\, ,
\ee
where the reduced Planck mass $M_{\mathrm{Planck}} = 1/ \kappa = 1/\sqrt{8\pi G}$ is 
a typical scale of the gravity.
Then, if we use the counter term to obtain the observed small value of the vacuum energy 
from the large theoretical one in (\ref{01B}),
we need a fine-tuning in the extremely unnatural way. 
We should note that the above fine-tuning is not so improved even if we introduce the supersymmetry, 
where the fermionic contributions to the vacuum energy are opposite to the bosonic contributions.
Because the supersymmetry is broken in a high energy region, we find 
\begin{align}
\rho_\mathrm{vacuum} 
=& \frac{1}{\left( 2 \pi \right)^3}\int d^3 k \left[ \frac{1}{2} \sqrt{ k^2 
+ m_{\mathrm{boson}}^2 } - \frac{1}{2} \sqrt{ k^2 + m_{\mathrm{fermion}}^2 } \right]
\nonumber \\ 
\sim & \Lambda_{\mathrm{cutoff}}^{2} \Lambda_{\cancel{\mathrm{SUSY}}}^{2}
\end{align}
where the scale of the supersymmetry breaking is defined by 
$\Lambda_{\cancel{\mathrm{SUSY}}}^{2} 
= m_{\mathrm{boson}}^2 - m_{\mathrm{fermion}}^2$, where $m_{\mathrm{boson}}$ 
is the mass of the bosonic mode and $m_{\mathrm{fermion}}$ is the mass of the fermionic 
mode. 
The fine-tuning problem of the cosmological constant, therefore, would imply the necessity 
of a new paradigm which includes the breakthrough beyond the standard knowledge based on the quantum field theory.

So far several models trying to solve the cosmological constant problem have been 
proposed\footnote{
For the discussion why the vacuum energy is so small, see 
\cite{Burgess:2013ara} for example. 
For the model using the topological field theory, see
\cite{Mori:2017dhe,Nojiri:2016mlb}
}  \cite{Anderson:1971pn,
Buchmuller:1988wx,Buchmuller:1988yn, Henneaux:1989zc,Unruh:1988in,
Ng:1990xz,Finkelstein:2000pg,Alvarez:2005iy,Alvarez:2006uu,
Abbassi:2007bq,Ellis:2010uc,Jain:2012cw,
Singh:2012sx,Kluson:2014esa,Padilla:2014yea,Barcelo:2014mua,
Barcelo:2014qva,Burger:2015kie,
Alvarez:2015sba,Jain:2012gc,Jain:2011jc,Cho:2014taa,Basak:2015swx,
Gao:2014nia,Eichhorn:2015bna,
Saltas:2014cta,Nojiri:2015sfd,Batra:2008cc,Shaw:2010pq,Barrow:2010xt,Carballo-Rubio:2015kaa}. 
In these models, an interesting mechanism, called ``sequestering'' mechanism proposed in \cite{Kaloper:2013zca,Kaloper:2014dqa,Kaloper:2015jra},
plays an important role to solve the fine-tuning problem. 
The remarkable feature of the sequestering mechanism is that two global variables are introduced.
The variations of the action with respect to these global variables lead to the constraint equations, 
where the quantum corrections coming from matters are explicitly canceled by the classical dynamics of gravity.
After the cancellation of quantum corrections, 
an effective cosmological constant remains in the equation of motion, 
and it is given by the matter density averaged with respect to the whole space-time, whose volume is given by  $\int d^{4}x \sqrt{-g}$.
Therefore, the model would naturally explain that the observed vacuum energy is tiny in the large and old Universe.

When one evaluates the value of the vacuum energy in the theories with the sequestering mechanism, 
it is significant to study whether one can obtain the appropriate cosmic history 
because the space-time average is literally written in terms of the four-dimensional volume of the Universe;
thus, the value of the effective cosmological constant depends on the cosmological model via the space-time average,
and the whole volume of the space-time should be finite in order to make the model well defined.

To realize such a model where the four-dimensional volume is finite, 
we study the scalar-tensor theory based on the formulation of the reconstruction in 
\cite{Nojiri:2005pu,Capozziello:2005tf}. 
We propose a model in which the Universe started from the big-bang
and, through the accelerated expansion corresponding to the present dark energy era,
the expansion becomes to be decelerated and turns to shrink to the big-crunch.
The four-dimensional volume in such a model is finite, and one can calculate 
the space-time average of the energy density from the matter fields.
And therefore, we obtain the effective vacuum energy corresponding to the cosmological constant 
and compare the observed value.

This paper is organized as follows.
First, we give a brief introduction of the sequestering mechanism in the general relativity
and see how the large quantum corrections are removed in the equation of motion 
in Sec.~\ref{SecII}.
We formulate the sequestering mechanism in the scalar-tensor theory
and review the formulation of the reconstruction in Sec.~\ref{SecIII}.
Finally, we evaluate the ratio of dark energy to the matters according to the specific 
cosmological evolution in Sec.~\ref{SecIV}.
We also study the parameter region in which the observed value of the dark energy is consistent.
 
\section{Sequestering Mechanism \label{SecII}}

We now review the sequestering mechanism in the general relativity 
\cite{Kaloper:2013zca,Kaloper:2014dqa,Kaloper:2015jra}.
We begin with the following action:
\begin{align}
S =& 
\int d^{4}x \ \sqrt{-g} 
\left[ \frac{1}{2\kappa^{2}} R - \Lambda
+ \lambda^{4}\mathcal{L}_\mathrm{m}(\lambda^{-2}g^{\mu\nu},\Psi)
\right]
+ \sigma\left(\frac{\Lambda}{\mu^{4}\lambda^{4}} \right) \, .
\label{eq:eq0}
\end{align}
Here $\mathcal{L}_\mathrm{m}$ denotes the Lagrangian density for the matter fields
which minimally couple with the metric $\tilde{g}_{\mu \nu} = \lambda^{2}g_{\mu \nu}$ 
and $\kappa^2 = 8\pi G$, where $G$ is the gravitational constant. 
The scalar curvature $R$ is constructed by $g_{\mu \nu}$. 
We should also note  
\begin{align}
\label{E6}
\sqrt{-g} \lambda^{4}\mathcal{L}_\mathrm{m} (\lambda^{-2}g^{\mu\nu},\Psi) 
= \sqrt{ - \tilde{g}} \mathcal{L}_\mathrm{m}(\tilde{g}^{\mu\nu},\Psi) \, .
\end{align}
The variables $\Lambda$ and $\lambda$ are global and they do not depend on the space-time 
coordinates $x$. 
And $\sigma$ is a differentiable function of the dimensionless combination 
of $\Lambda$ and $\lambda$ with the mass scale parameter $\mu$ introduced by the 
dimensional reasons.
Note that the global variable $\lambda$ is responsible for the hierarchy between the typical 
matter scale and the Planck scale.
For a scalar field $\phi$, the field redefinition related to scaling of 
the metric $\tilde{g}_{\mu \nu} = \lambda^2 g_{\mu \nu}$ leads to
\begin{align}
\label{E7}
\sqrt{-\tilde{g}} \left[ - \frac{1}{2} \tilde{g}^{\mu \nu} \partial_{\mu} \phi 
\partial_{\nu} \phi -  \frac{1}{2} m^{2} \phi^2 \right]
= \sqrt{-g} \left[ - \frac{1}{2} \tilde{g}^{\mu \nu} \partial_{\mu} \Phi 
\partial_{\nu} \Phi -  \frac{1}{2} m^{2}_{\mathrm{phys}} \Phi^{2}  \right] \, ,
\end{align}
where $\Phi = \lambda \phi$ and the physical mass is defined by 
$m_{\mathrm{phys}} = \lambda m$,
thus we find 
\begin{align}
\label{E8}
\frac{m_{\mathrm{phys}}}{M_{\mathrm{Planck}}} 
= \lambda \frac{m}{M_{\mathrm{Planck}}} \, .
\end{align}
By the variation of the action with respect to $\delta g^{\mu\nu}$,
one finds
\begin{align}
\label{E9}
\delta_{g}S 
=& 
\int d^{4}x \ 
\left[ \sqrt{-g}
\left\{ \frac{1}{2\kappa^{2}} G_{\mu \nu} + \frac{1}{2} \Lambda  g_{\mu\nu} 
\right \} \delta g^{\mu \nu} 
- \frac{1}{2} \sqrt{-\tilde{g}} \tilde{T}_{\mu \nu} \delta \tilde{g}^{\mu \nu}
\right]
\nonumber \\
=&
\int d^{4}x \ 
\sqrt{-g} \left[ 
\frac{1}{2\kappa^{2}} G_{\mu \nu} + \frac{1}{2} \Lambda g_{\mu\nu}
- \frac{1}{2} \lambda^{2} \tilde{T}_{\mu \nu}  \right] \delta g^{\mu \nu} \, ,
\end{align}
where we used $\delta \tilde{g}^{\mu \nu} = \lambda^{-2} \delta g^{\mu \nu}$ 
and $G_{\mu \nu}$ is the Einstein tensor, $G_{\mu \nu}=R_{\mu\nu} - \frac{1}{2}g_{\mu\nu} R$.
The energy-momentum tensor $\tilde{T}_{\mu \nu}$ is defined as
\begin{align}
\label{E10}
\tilde{T}_{\mu\nu} (\tilde{g}^{\mu \nu}, \Psi)
\equiv
\frac{-2}{\sqrt{-\tilde{g}}}
\frac{\delta\left(\sqrt{-\tilde{g}}{\cal L}_\mathrm{m}
\left(\tilde{g}^{\mu\nu},\Psi \right) \right)}{\delta\tilde{g}^{\mu\nu}} \, .
\end{align}
We should note that we may define another energy-momentum tensor $T_{\mu \nu}$ according 
to the variation with respect to the metric $g_{\mu \nu}$.
The relation between the two energy-momentum tensors, $T_{\mu \nu}$ and 
$\tilde{T}_{\mu \nu}$, is given by
\begin{align}
T_{\mu \nu} (\lambda^{-2}g^{\mu \nu}, \Psi)
=& 
\frac{- 2}{\sqrt{-g}} 
\frac{ \delta \left( \sqrt{-g}  \lambda^{4} \mathcal{L}_\mathrm{m}
(\lambda^{-2} g^{\mu \nu}, \Psi) \right)}{\delta g^{\mu \nu}}
\nonumber \\
=& 
\frac{- 2}{ \lambda^{-4} \sqrt{-\tilde{g}}} 
\frac{  \delta \left( \sqrt{-\tilde{g}} \mathcal{L}_\mathrm{m}
(\tilde{g}^{\mu \nu}, \Psi) \right)}{ \lambda^{2} \delta \tilde{g}^{\mu \nu}}
\nonumber \\
=&
\lambda^{2} \tilde{T}_{\mu \nu}  (\tilde{g}^{\mu \nu}, \Psi) \, .
\label{relation_btw_energy-momentum_tensor}
\end{align}
We finally obtain the equation of motion for the metric $g_{\mu \nu}$ as follows, 
\begin{align}
\frac{1}{\kappa^{2}} G_{\mu \nu} + \Lambda g_{\mu \nu} = T_{\mu \nu} \, .
\label{equation_of_motion1}
\end{align}
In addition to the equation of motion for $g_{\mu \nu}$, 
the variations with respect to $\Lambda$ and $\lambda$ give the constraint equations 
as follows, respectively,
\begin{align}
\frac{\sigma^{\prime}}{\lambda^{4}\mu^{4}} 
=& \int d^{4}x \sqrt{-g} \, , 
\label{constraint1} \\
4 \Lambda \frac{\sigma^{\prime}}{\lambda^{4}\mu^{4}} 
=&
\int d^{4}x \sqrt{-g} T^{\mu}_{\ \mu} \, ,
\label{constraint2}
\end{align}
where we have used
\begin{align}
\label{E15}
\delta_{\lambda} \left( 
\sqrt{-g} \lambda^{4}\mathcal{L}_\mathrm{m}(\lambda^{-2}g^{\mu\nu},\Psi)
\right)
=&
\delta_{\lambda} \left(
\sqrt{-\tilde{g}} \mathcal{L}_\mathrm{m}(\tilde{g}^{\mu\nu},\Psi)
\right)
\nonumber \\
=&
- 2\lambda^{-1} \tilde{g}^{\mu \nu} 
\delta_{\tilde{g}} \left(
\sqrt{-\tilde{g}} \mathcal{L}_\mathrm{m}(\tilde{g}^{\mu\nu},\Psi)
\right)
\nonumber \\
=&
\lambda^{-1} \sqrt{-\tilde{g}} \tilde{T}^{\mu}_{\ \mu} \, .
\end{align}

By dividing Eq.~(\ref{constraint2}) by Eq.~(\ref{constraint1}) in both sides,  
one finds that the global variable $\Lambda$ can be expressed in terms of 
the energy-momentum tensor,
\begin{align}
\Lambda = \frac{1}{4} \langle T^{\mu}_{\ \mu} \rangle \, ,
\label{lambda_average}
\end{align}
where $\langle \mathcal{O} \rangle$ is four-dimensional space-time
volume average of the quantity $\mathcal{O}$, 
defined as follows:
\begin{align}
\langle \mathcal{O} \rangle 
= \frac{\int d^{4}x \sqrt{-g} \, \mathcal{O}}{\int d^{4}x \sqrt{-g}} \, .
\label{4-volume_average}
\end{align}
Note that, strictly speaking, the global average is well defined 
if the space-time volume $\int d^{4}x\ \sqrt{-g}$ is finite. 
Substituting Eq.~(\ref{lambda_average}) into Eq.~(\ref{equation_of_motion1}),
the equation of motion for the metric $g_{\mu \nu}$ has the following form,
\begin{align}
\frac{1}{\kappa^{2}} G_{\mu \nu} 
= - \frac{1}{4} \langle T^{\mu}_{\ \mu} \rangle g_{\mu \nu} + T_{\mu \nu} \, .
\label{eom1}
\end{align}

Next, we divide the matter Lagrangian into two parts and extract the vacuum energy 
obtained from the quantum corrections of matter fields:
\begin{align}
\label{E19}
\sqrt{-g} \lambda^{4} \mathcal{L}(\lambda^{-2}g^{\mu \nu}, \Psi) 
= \sqrt{-g} \left[ -  V_{\mathrm{vac}} + \lambda^{4} \Delta \mathcal{L}_{\mathrm{eff}} (\lambda^{-2}g^{\mu \nu}, \Psi) \right] \, ,
\end{align}
or equivalently 
\begin{align}
\label{E20}
\sqrt{-\tilde{g}} \mathcal{L}(\tilde{g}^{\mu \nu}, \Psi)
= \sqrt{-\tilde{g}} \left[ - \lambda^{-4} V_{\mathrm{vac}} 
+ \Delta \mathcal{L}_{\mathrm{eff}} (\tilde{g}^{\mu \nu}, \Psi) \right] \, .
\end{align}
Then, the corresponding energy-momentum tensor follows as
\begin{align}
\label{E21}
\tilde{T}_{\mu\nu} (\tilde{g}^{\mu \nu}, \Psi)
=& - \lambda^{-4} V_{\mathrm{vac}} \tilde{g}_{\mu \nu} + \tilde{\tau}_{\mu \nu} \, ,
\end{align}
where $\tau_{\mu \nu}$ expresses the energy-momentum tensor of the matter field 
in which the vacuum energy is subtracted:
\begin{align}
\label{E22}
\tilde{\tau}_{\mu \nu}
= \frac{-2}{\sqrt{-\tilde{g}}}
\frac{\delta \left(\sqrt{-\tilde{g}} \Delta \mathcal{L}_{\mathrm{eff}} 
(\tilde{g}^{\mu \nu}, \Psi)  \right)}{\delta\tilde{g}^{\mu\nu}} \, .
\end{align}
According to Eq.~(\ref{relation_btw_energy-momentum_tensor}), 
which gives us the relation between 
the energy-momentum tensors obtained by the variation with respect to $\tilde{g}_{\mu \nu}$ 
and $g_{\mu \nu}$, the energy-momentum tensor $T_{\mu \nu}$ is expressed as
\begin{align}
T_{\mu \nu} = - V_{\mathrm{vac}} g_{\mu \nu} + \tau_{\mu \nu} \, .
\label{separated_energy_momentum_tensor}
\end{align}
Finally, Eq.~(\ref{eom1}) is given by the following form:
\begin{align}
\frac{1}{\kappa^{2}} G_{\mu \nu} 
=&
- \frac{1}{4} \langle \tau^{\mu}_{\ \mu} \rangle  g_{\mu \nu} + \tau_{\mu \nu} \, .
\end{align}

One finds that there is a residual effective cosmological constant coming from the space-time average 
of the trace of matter fields:
\begin{align}
G_{\mu \nu} + \Lambda_{\mathrm{eff}} g_{\mu \nu} 
= \kappa^{2} \tau_{\mu \nu} \, ,
\label{eom2}
\end{align}
where 
\begin{align}
\label{lambda_eff}
\Lambda_{\mathrm{eff}} = \frac{\kappa^{2}}{4} \langle \tau^{\mu}_{\ \mu} \rangle \, .
\end{align}
Thus, the vacuum energy from the quantum corrections of matter fields is canceled 
in the equation of motion Eq.~(\ref{eom2}),
and that the numerical value of the residual constant $\Lambda_{\mathrm{eff}}$ is automatically 
small if our Universe is large enough and old.

We should note that the constraint equations (\ref{constraint1}) and (\ref{constraint2}) 
give us
\begin{align}
\label{E27}
\frac{\Lambda}{\lambda^{4}\mu^{4}} 
= \frac{1}{4\mu^{4}} \langle \tilde{T}^{\mu}_{\ \mu} \rangle \, ,
\end{align}
which leads to
\begin{align}
\label{E28}
\lambda 
=& \left[ \frac{\sigma^{\prime}\left( \frac{1}{4\mu^{4}} 
\langle \tilde{T}^{\mu}_{\ \mu} \rangle  \right)}{\mu^{4} \int d^{4}x \sqrt{-g} }\right]^{1/4}
\, , \qquad 
\Lambda 
= \frac{\sigma^{\prime}\left( \frac{1}{4\mu^{4}} \langle \tilde{T}^{\mu}_{\ \mu} 
\rangle  \right) \frac{1}{4\mu^{2}} 
\langle \tilde{T}^{\mu}_{\ \mu} \rangle}{\int d^{4}x \sqrt{-g}}\, .
\end{align}
We find that the space-time average relates to the hierarchy between the particle mass scale 
and the Planck mass scale, which is sensitive to the choice of $\sigma$.
The form of function $\sigma$ is, therefore, not arbitrary but would be determined or constrained by 
the phenomenological requirements.
Furthermore, $\Lambda$ and $\lambda$ are given as functions of the four-dimensional volume 
$\int d^{4}x \sqrt{-g}$, and the space-time volume of the Universe is the independent variable in this theory.

We also mention about the two symmetries in the sequestering mechanism which ensures 
the cancellation of quantum corrections in the vacuum energy.
First, we find the scale invariance in the action.
Under the following scale transformation,
\begin{align}
\label{E29}
\lambda \rightarrow \Omega \lambda \, , \quad g_{\mu \nu} \rightarrow 
\Omega^{-2}g_{\mu \nu} \, , \ \quad  \Lambda \rightarrow \Omega^{4} \Lambda \, , 
\end{align}
the action changes by
\begin{align}
\label{E30}
S \to S_\Omega \equiv & \frac{1}{2\kappa^{2}} \Omega^{-2} \int d^{4}x \sqrt{-g}R 
\nonumber \\
=& \frac{1}{2\kappa^{2}} \Omega^{-2} \langle R \rangle \int d^{4}x \sqrt{-g} 
\, .
\end{align}
In fact, the scaling symmetry is broken by the gravity sector and 
the symmetry is approximate one.
This symmetry breaking is, however, generated by the mediation from the gravitational sector 
through the $\int d^{4}x \sqrt{-g}$, and therefore the breaking is weak.
Furthermore, we can find that the action is exactly invariant under the scale transformation on shell:
The trace of the equation of motion (\ref{eom1}) leads to
\begin{align}
\label{E31}
\frac{1}{\kappa^2}G^{\mu}_{\ \mu} = T^{\mu}_{\ \mu} - \langle T^{\mu}_{\ \nu} \rangle \, ,
\end{align}
and, by taking the space-time average for both hand sides, we find $\langle R \rangle = 0$.

Second symmetry is the invariance under the following shift transformation,
\begin{align}
\label{E32}
\mathcal{L} \rightarrow \mathcal{L} + \epsilon m^{4} \, , \quad 
\Lambda \rightarrow \Lambda - \epsilon \lambda^{4}m^{4} \, .
\end{align}
Under the transformation, the variation of the action is given by 
\begin{align}
\label{E33}
\delta S 
=& \sigma \left( \frac{\Lambda}{\lambda^{4}\mu^{4}} 
 - \epsilon \frac{m^{4}}{\mu^{4}} \right) 
 - \sigma \left( \frac{\Lambda}{\lambda^{4}\mu^{4}}  \right)
\nonumber \\
\approx & - \epsilon \sigma^{\prime} \frac{m^{4}}{\mu^{4}} \, .
\end{align}
The shift symmetry is broken as well as the scale symmetry, but the breaking is also weak because
\begin{align}
\label{E34}
\delta S 
\approx& - \epsilon m^{4} \lambda^{4} \cdot \frac{\sigma^{\prime}}{\lambda^{4} \mu^{4}}
\nonumber \\
=& - \epsilon \left( \frac{m_{\mathrm{phys}}}{M_{\mathrm{Planck}}} \right)^{4} 
M_{\mathrm{Planck}}^{4} \int d^{4}x \sqrt{-g} \, ,
\end{align}
which is small when $m_{\mathrm{phys}}/M_{\mathrm{Planck}} \ll 1$.
By using the constraints and equation of motion, we find that the shift can be absorbed 
in the redefinitions of  the global variables although the metric is not changed.

These two symmetries are the key to understand how the sequestering mechanism works.
The scaling symmetry ensures that the vacuum energy at an arbitrary order has the same couplings
with the gravity as the classical one,
and the shift symmetry removes the vacuum energy from the Lagrangian.
Thus, the quantum corrections at all orders are canceled without any tuning in order by order. 

In the general relativity, the sequestering mechanism does not give a positive cosmological constant unless we consider the matter sector which causes the deceleration expansion.
For example, in a nonrelativistic perfect fluid,
the trace of matter field is given by $\tau^{\mu}_{\mu}=-\rho$.
Because we assume that the four-dimensional volume and the energy density is positive,
$\Lambda_{\mathrm{eff}}$, (\ref{lambda_eff}), become negative, and it corresponds to the negative cosmological constant.
A negative cosmological constant does not give the acceleration expansion era.
So we need to introduce the candidate of dark energy.
In Sec.~\ref{SecIII}, we will use scalar-tensor theory, which is one of the modified gravity, to realize the expanding Universe. 

\section{Scalar-Tensor Theory with Sequestering Mechanism \label{SecIII}}

\subsection{Action and equations of motion}

In the previous section, 
we introduced the basic property of the sequestering mechanism.
Here, we should note that the sequestering mechanism to remove the large vacuum energy can 
be used in more general frameworks because this mechanism does not depend on the gravitational theory itself.
In this paper, we consider a particular time-evolution of the Universe by using the reconstruction method known in the scalar-tensor theory,
and we perform the space-time average for a perfect fluid as the baryon and dark matter.

We consider the following model with a scalar field $\phi$, 
\begin{equation}
\label{eq:050101}
S = \int d^{4}x \ \sqrt{-g} \left[ \frac{1}{2\kappa^2}R 
 - h(\phi)\left(\nabla\phi \right)^{2} - V(\phi) - \Lambda
+ \lambda^{4} \mathcal{L}_\mathrm{m}(\lambda^{-2}g^{\mu\nu},\Psi) \right]
+ \sigma\left(\frac{\Lambda}{\mu^{4}\lambda^{4}} \right) \, .
\end{equation}
Without the contributions from the scalar field $\phi$, that is, 
$h(\phi)=V(\phi)=0$, the model in (\ref{eq:050101}) reduces to 
the action of the original sequestering model Eq.~(\ref{eq:eq0}) proposed in 
\cite{Kaloper:2013zca,Kaloper:2014dqa,Kaloper:2015jra}.

By the variation of the action with respect to the metric $g_{\mu\nu}$, 
we obtain 
\be
\label{eq:050102}
\frac{1}{2\kappa^2}G_{\mu\nu} - h\nabla_{\mu}\phi \nabla_{\nu}\phi
+ \frac{1}{2}h\left(\nabla\phi \right)^{2}g_{\mu\nu}
+ \frac{1}{2}\left(V + \Lambda \right)g_{\mu\nu} - \frac{1}{2} T_{\mu\nu}
= 0\, .
\ee
The variations with respect to $\Lambda$ and $\lambda$ give us the same constraints as 
in  Eqs.~(\ref{constraint1}) and (\ref{constraint2}).
Substituting these constraints into Eq.~(\ref{eq:050102}),  we obtain 
\be
\label{eq:050106}
\frac{1}{2\kappa^2}G^{\mu}_{\ \nu}
 - h\nabla^{\mu}\phi \nabla_{\nu}\phi
+ \frac{1}{2}h\left(\nabla\phi \right)^{2}\delta^{\mu}_{\ \nu}
+ \frac{1}{2}V\delta^{\mu}_{\ \nu}
+ \frac{1}{4}\left\langle 
\frac{1}{2}T^{\alpha}_{\ \alpha} \right\rangle\delta^{\mu}_{\ \nu}
 - \frac{1}{2}T^{\mu}_{\ \nu}
= 0 \, .
\ee
Decomposing the energy-momentum tensor into the sum of the vacuum energy and others 
as in Eq.~(\ref{separated_energy_momentum_tensor}),
we find that Eq.~(\ref{eq:050106}) can be expressed as 
\be
\label{eq:050109}
\frac{1}{2\kappa^2}G^{\mu}_{\ \nu} - h\nabla^{\mu}\phi \nabla_{\nu}\phi
+ \frac{1}{2}h\left(\nabla\phi \right)^{2}\delta^{\mu}_{\ \nu}
+ \frac{1}{2}V\delta^{\mu}_{\ \nu}
+ \frac{1}{4}\left\langle \frac{1}{2}\tau^{\alpha}_{\ \alpha} 
\right\rangle\delta^{\mu}_{\ \nu} - \frac{1}{2}\tau^{\mu}_{\ \nu}
= 0\, .
\ee
Defining $\Lambda_\mathrm{eff}$ by  
\be
\label{eq:040115}
\Lambda_\mathrm{eff}
= \frac{\kappa^2}{4} \left\langle \tau^{\alpha}_{\ \alpha} \right\rangle\, ,
\ee
we further rewrite the Eq.~(\ref{eq:050109}) as 
\be
\label{eq:050110}
\frac{1}{2\kappa^2}G^{\mu}_{\ \nu} - h\nabla^{\mu}\phi \nabla_{\nu}\phi
+ \frac{1}{2}h\left(\nabla\phi \right)^{2}\delta^{\mu}_{\ \nu}
+ \frac{1}{2}V\delta^{\mu}_{\ \nu}
+ \frac{1}{2\kappa^2} \Lambda_\mathrm{eff} \delta^{\mu}_{\ \nu}
- \frac{1}{2}\tau^{\mu}_{\ \nu} = 0 \, .
\ee

\subsection{Cosmological solution}

By using the above equation (\ref{eq:050110}) in the FRW metric, we investigate the time-evolution of the Universe. 
We assume the FRW metric as follows, 
\be
\label{eq:050111}
ds^{2} = - dt^{2} + a(t)^{2}\gamma_{ij}dx^{i}dx^{j}
= - dt^{2} + a^{2}
\left( \frac{1}{1-Kr^{2}}dr^{2}  + r^{2}d\theta^{2} 
+ r^{2}\sin^{2}\theta d\varphi^{2} \right) \, ,
\ee
and we only consider the closed Universe where the curvature of the space $K > 0$ because 
we require the volume of the space-time should be finite. 

We now assume $\tau^{\mu}_{\ \nu}$ is given by perfect fluid, 
\be
\label{eq:050112}
\tau^{\mu}_{\ \nu} = \mathrm{diag}(-\rho, p, p, p) \, ,
\ee
and the scalar field $\phi$ only depends on the cosmological time, 
\be
\label{eq:050117}
\phi = \phi(t) \, .
\ee
Then the $(0,0)$ component of (\ref{eq:050110}) is given by 
\be
\label{eq:050118}
H^{2} = \frac{\kappa^2}{3}\rho - \frac{K}{a^{2}} 
+ \frac{2\kappa^2}{3}\left(
\frac{1}{2}h{\dot \phi}^{2}
+ \frac{1}{2}V + \frac{1}{2\kappa^2}\Lambda_\mathrm{eff} \right) \, ,
\ee
and $(i, i)$ components are 
\be
\label{eq:050119}
3H^{2} + 2\dot{H} = -\kappa^2 p - \frac{K}{a^{2}} 
+ 2\kappa^2\left(
-\frac{1}{2}h{\dot \phi}^{2}
+ \frac{1}{2}V 
+ \frac{1}{2\kappa^2}\Lambda_\mathrm{eff} \right) \, .
\ee
Other components become identities. 
By combining (\ref{eq:050118}) and (\ref{eq:050119}), we find 
\begin{align}
\label{eq:050122}
V\left(\phi\right) =& \frac{3}{\kappa^2}H^{2}
+ \frac{1}{\kappa^2} \dot{H} - \frac{1}{2}\left(
\rho_\mathrm{m} - p_\mathrm{m} \right) - \frac{1}{\kappa^2} \Lambda_\mathrm{eff}
+ \frac{2}{\kappa^2}\frac{K}{a^{2}} \, , \\
\label{eq:050123}
h\left(\phi\right)\dot{\phi}^{2} =& - \frac{1}{\kappa^2}\dot{H} 
 - \frac{1}{2}\left( \rho_\mathrm{m} + p_\mathrm{m} \right)
+\frac{1}{\kappa^2}\frac{K}{a^{2}}\, .
\end{align}
Let $f(\phi)$ is a function of the scalar field $\phi$. 
If the potential $V(\phi)$ and the kinetic function $h(\phi)$ are given 
in terms of $f(\phi)$ as follows, 
\begin{align}
\label{eq:050123-24-1}
V\left(\phi\right) =& \frac{3}{\kappa^2} f(\phi)^{2}
+ \frac{1}{\kappa^2} f^{\prime}(\phi) - \frac{1}{2}\rho_\mathrm{m} \left( t = \phi \right)
 - \frac{1}{4}\left\langle \tau^{\alpha}_{\ \alpha} \right\rangle
+ \frac{2}{\kappa^2}\frac{K}{a \left( t = \phi \right)^{2}} \, , \nn
h\left(\phi\right) =& - \frac{1}{\kappa^2} f^{\prime}(\phi)
 - \frac{1}{2}\rho_\mathrm{m}  \left( t = \phi \right) 
+ \frac{1}{\kappa^2}\frac{K}{a \left( t = \phi \right)^{2}} \, .
\end{align}
Then the solution of Eqs.~(\ref{eq:050122}) and (\ref{eq:050123}) 
and therefore the solution of (\ref{eq:050118}) and (\ref{eq:050119}) 
is given by 
\be
\label{eq:050124}
H = f\left(\phi=t \right) \, , \quad \phi = t\, .
\ee
Therefore any evolution of the expansion in the Universe given by 
the function $H=f(t)$ can be realized by choosing $V\left(\phi\right)$ 
and $h\left(\phi\right)$ as in (\ref{eq:050123-24-1}). 
We should note that if $h \left( \phi \right)$ is negative, the scalar field becomes ghost which generates the negative norm states in the quantum theory and therefore inconsistent. 

\section{A Concrete Model of Sequestering Mechanism \label{SecIV}}

As we have mentioned, we like to have a model where the volume of the 
space-time $\int d^{4}x\ \sqrt{-g}$ is finite in order that the 
global average $\left< \mathcal{O} \right>$ in Eq.~(\ref{4-volume_average}) for relevant physical 
operator $\mathcal{O}$ should be well defined. 
As we have shown in the last section, arbitrary evolution of the expansion 
in the Universe can be realized by choosing the potential $V\left(\phi\right)$ 
and the kinetic factor $h\left(\phi\right)$ to satisfy the equations 
in (\ref{eq:050123-24-1}) as in the formulation of the reconstruction 
in \cite{Nojiri:2005pu,Capozziello:2005tf}. 
Then in this section, we construct a model where 
the curvature of the space $K>0$ and the Universe acceleratingly expands in the late time 
(for the review about the modified gravity theories related to the accelerating expansion 
of the Universe, see \cite{Nojiri:2006ri,Nojiri:2010wj,Nojiri:2017ncd,Capozziello:2011et}).
We also require that the Universe finally turns to shrink, and therefore, 
the whole volume of the space-time is finite. 
 
We consider the model, where the scale factor is given by 
\be
\label{eq:050401}
a(t) = \left\{a_{1}(t) \right\}^{1/n}\, , \quad 
a_{1}(t) \equiv \alpha\left(\frac{1}{12}t^{4} - \frac{1}{2}t_{1}^{2}t^{2} 
+ C \right) \, , \quad 
C \equiv -\frac{1}{12}t_{0}^{4} + \frac{1}{2}t_{1}^{2}t_{0}^{2} \, .
\ee
Here $t_0$ and $t_1$ are positive constants. 
When $t=\pm t_0$, we find $a\left( \pm t_0 \right)=0$ and therefore 
$t=-t_0$ corresponds to the big-bang and $t=t_0$ to the big-crunch. 
The scale factor (\ref{eq:050401}) is symmetric under the reflection of 
the time $t\to -t$ and the expanding Universe turns to shrink at $t=0$. 

First we now check the signature of $h(\phi)$ in (\ref{eq:050123-24-1}). 
As we have mentioned, if $h(\phi=t)$ becomes negative, there appears the ghost, and the theory becomes inconsistent. 
Now $h(\phi)$ is explicitly given by
\begin{align}
\label{eq:050434}
h(t) =& \frac{H_{0}^{2}}{\kappa} a(t)^{-3} \left[
\frac{\alpha^{2}}{36nH_{0}^{2}} \left\{a_{1}(t) \right\}^{\frac{3}{n} -2} 
\left\{G_{1}\left(X \right) + 36Ct_{1}^{2} \right\}
 -\frac{3}{2}\Omega_{m0} - \Omega_{K0}\left\{a_{1}(t) \right\}^{\frac{1}{n}} 
\right] \nn
 =& \frac{H_{0}^{2}}{\kappa} a(t)^{\frac{2}{n}-5} \left[
\frac{\alpha^{2}}{36nH_{0}^{2}} 
\left\{G_{1}\left(X \right) + 36Ct_{1}^{2} \right\}
 -\frac{3}{2}\Omega_{m0}\left\{a_{1}(t) \right\}^{- \frac{3}{n} +2} 
 - \Omega_{K0}\left\{a_{1}(t) \right\}^{- \frac{2}{n} + 2} \right] \, .
\end{align}
Here $X\equiv t^2$, $H_{0}$ is the present value of the Hubble rate $H$, 
$\Omega_{m0}$ and $\Omega_{K0}$ are the values of the present density 
parameters of the dust and the curvature, and 
\begin{align}
\label{eq:050419.20.1}
a_{1}(t) \equiv& \frac{\alpha t_{1}^{4}}{12y^{2}} 
\left\{1-\left(\frac{t}{t_{0}} \right)^{2} \right\}\left\{
6y-1-\left(\frac{t}{t_{0}} \right)^{2} \right\}\, , 
\quad y\equiv \left(\frac{t_{1}}{t_{0}}\right)^{2} \quad \left(0 < y < 1 \right)
\, , \nn
G_{1}\left(X \right) \equiv&  X\left(X^{2} - A_{3}X + A_{4} \right) \, , \quad 
A_{3} \equiv 3t_{1}^{2} \, , \quad 
A_{4} \equiv 36t_{1}^{4}\left(\frac{1}{2} - \frac{C}{t_{1}^{4}} \right) \, .
\end{align}
We now investigate the behavior of $h(\phi=t)$ when $t\to t_0$. 
If $n<\frac{3}{2}$, because $a_1(t) \to 0$ when $t\to t_0$, the second or the 
third term in the r.h.s. of (\ref{eq:050434}) and therefore $h(\phi=t)$ becomes 
negative and therefore there appears the ghost. 
Then we require $n\ \geq\ \frac{3}{2}$. 
When $n>\frac{3}{2}$, in (\ref{eq:050434}), the 
first term dominates and therefore we obtain the following condition, 
\be
\label{eq:050435}
G_{1}\left(X=t_{0}^{2} \right) + 36Ct_{1}^{2} > 0 \, .
\ee
The left-hand side of Eq.~(\ref{eq:050435}) has the following form, 
\be
\label{eq:050437}
G_{1}\left(X=t_{0}^{2} \right) + 36Ct_{1}^{2}
=\frac{4t_{1}^{6}}{y^{3}}\left(3y-1 \right)^{2} \, ,
\ee
and therefore the condition (\ref{eq:050435}) is always satisfied when 
$t\to t_0$. 
We also need to consider the signature of $h(\phi=t)$ for general $t$ in $-t_0 < t < t_0$. 
We should  note $G_1 \left( X \right)$ has minimum at $X=X_+$ for $X=t^2 >0$, where 
\be
\label{Xpm}
X_\pm \equiv \frac{A_3 \pm \sqrt{A_3^2 - 3 A_4 }}{3}\, .
\ee
Then because 
\begin{align}
\label{hp1}
& \frac{\alpha^{2}}{36nH_{0}^{2}} 
\left\{G_{1}\left(X \right) + 36Ct_{1}^{2} \right\}
 -\frac{3}{2}\Omega_{m0}\left\{a_{1}(t) \right\}^{- \frac{3}{n} +2} 
 - \Omega_{K0}\left\{a_{1}(t) \right\}^{- \frac{2}{n} + 2} \nn
> & \frac{\alpha^{2}}{36nH_{0}^{2}} 
\left\{G_{1}\left(X_+ \right) + 36Ct_{1}^{2} \right\}
 -\frac{3}{2}\Omega_{m0}\left\{a_{1}(0) \right\}^{- \frac{3}{n} +2} 
 - \Omega_{K0}\left\{a_{1}(\pm t_{0}) \right\}^{- \frac{2}{n} + 2} \nn
= & \frac{\alpha^{2}}{36nH_{0}^{2}} 
\left\{G_{1}\left(X_+ \right) + 36Ct_{1}^{2} \right\}
 -\frac{3}{2}\Omega_{m0}\left\{ \alpha C \right\}^{- \frac{3}{n} +2} \, ,
\end{align}
the sufficient condition that $h(\phi=t)>0$ is given by 
\be
\label{hp2}
\frac{\alpha^{2}}{36nH_{0}^{2}} 
\left\{G_{1}\left(X_+ \right) + 36Ct_{1}^{2} \right\}
 -\frac{3}{2}\Omega_{m0}\left\{ \alpha C \right\}^{- \frac{3}{n} +2} > 0 \, .
\ee
Since $\alpha$ is obtained from the renormalization of the scale factor  for the present time $t_{p}$, 
(\ref{hp1}) is given by 
\begin{align}
\label{hp3}
&\frac{\alpha^{2}}{36nH_{0}^{2}} 
\left\{G_{1}\left(X_+ \right) + 36Ct_{1}^{2} \right\}
 -\frac{3}{2}\Omega_{m0}\left\{ \alpha C \right\}^{- \frac{3}{n} +2} \nn 
=& \frac{3z\Omega_{m0}}{2}\left(\frac{6y-1}{6(1-\epsilon^{2})y - (1-\epsilon^{4})} \right)^{2} 
\times \left[\frac{1}{6y-1}\left\{6(1-\epsilon^{2})y- (1-\epsilon^{4}) \right\} 
\right]^{\frac{3}{n}} 
\times \left[D_{n}(y,\epsilon)-\frac{1}{z} \right] \, ,
\end{align}
where the function $D_{n}\left(y,\epsilon \right)$ is defined as 
\begin{align}
\label{hpd}
D_{n}\left(y,\epsilon \right) 
\equiv \frac{16}{3n\Omega_{m0}}\frac{y^{3}}{(6y-1)^{2}}
\left[\frac{1}{6y-1}\left\{6(1-\epsilon^{2})y- (1-\epsilon^{4}) \right\} 
\right]^{-\frac{3}{n}} \times \left[8- \left\{\frac{1}{y^{2}}(1-y)(5y-1) 
\right\}^{\frac{3}{2}} \right] \, .
\end{align}
Here $\epsilon$ is defined by using the 
present time $t_p$ as $t_{p}=\epsilon t_{0}$ $\left(0<\epsilon<1\right)$. 
And $z$ is defined by $z \equiv \frac{1}{t_{0}^{2}H_{0}^{2}}$. 
 From (\ref{hp3}), we alter the sign of (\ref{hp1}) to the value of 
$D_{n}\left(y,\epsilon \right)$ versus $\frac{1}{z}$. 
Satisfying the condition (\ref{hp2}), we need to choose the parameter region 
$\left(z,n,y,\epsilon \right)$ of $D_{n}\left(y,\epsilon \right) > \frac{1}{z}$.
%}

The second derivative of $a(t)$ with respect 
the time $t$ is given by 
\be
\label{eq:050405}
\ddot{a}(t) = 
\frac{\alpha^{2}}{9n}\left\{a_{1}(t) \right\}^{1/n -2}\left[
F_{1}\left(X \right) -9Ct_{1}^{2} \right] \, , \quad 
F_{1}\left(X \right) \equiv \frac{4-n}{4n}X\left(X^{2} - A_{1}X + A_{2}
\right) \, .
\ee
Here
\be
\label{A12}
A_{1}\equiv \frac{8-n}{4-n}3t_{1}^{2}\, , \quad 
A_{2}\equiv \frac{4n}{4-n}\left(\frac{2-n}{2n} 
+ \frac{C}{t_{1}^{4}} \right)9t_{1}^{4} \, .
\ee
Because there is a singularity at $n=4$ in the expressions in (\ref{A12}), 
we restrict a constant $n$ to be 
\be
\label{nrange}
\frac{3}{2}<n<4\, .
\ee
When $n=\frac{3}{2}$, we also need to require, 
\be
\label{nrange2}
\frac{\alpha^{2}}{36nH_{0}^{2}}
\left\{G_{1}\left(X \right) + 36Ct_{1}^{2} \right\}
 -\frac{3}{2}\Omega_{m0} > 0 \, .
\ee
Because $\ddot a(t)<0$ at $X=t^2=0$ when the Universe starts to shrink, 
we require
\be
\label{C}
C =  -\frac{1}{12}t_{0}^{4} + \frac{1}{2}t_{1}^{2}t_{0}^{2} > 0\, .
\ee
On the other hand, $a_{1}(t)$ should be always positive in $t^{2} < t_{0}^{2}$. 
We now rewrite $a_{1}(t)$ as follows, 
\be
\label{eq:050419.20.1B}
a_{1} = \frac{\alpha t_{1}^{4}}{12y^{2}} 
\left\{1-\left(\frac{t}{t_{0}} \right)^{2} \right\}\left\{
6y-1-\left(\frac{t}{t_{0}} \right)^{2} \right\}\, .
\ee
Because $t^{2}/t_{0}^{2} < 1$, we find $6y-2 \geq 0$ and therefore
\be
\label{eq:050419.20.2}
\frac{1}{3} \leq y<1\, .
\ee
In terms of $y$, $C$ can be written as 
\be
\label{eq:050418}
\frac{C}{t_{1}^{4}} = \frac{1}{12 y^{2}}\left(6y-1 \right) \, .
\ee
The r.h.s. has a local maximum at $y=\frac{1}{3}$ and when $y>\frac{1}{3}$, 
the r.h.s. is a monotonically decreasing function of $y$. 
Therefore by using (\ref{eq:050419.20.2}), we find 
\be
\label{Crange}
\frac{5}{12}< \frac{C}{t_{1}^{4}} \leq \frac{3}{4} \, .
\ee
In order that the decelerating Universe turns to accelerate after the 
big-bang, 
and then, the accelerating Universe turns to decelerate again, 
$\ddot a(t)$ should vanish twice 
when $0<t^2=X<t_0^2$. 
Because 
\be
\label{eq:050407}
\frac{dF_{1}\left(X \right)}{dt}
= \frac{dX}{dt} \frac{3(4-n)}{4n}\left\{ \left(X-\frac{1}{3}A_{1} \right)^{2}
+\frac{1}{9}\left(3A_{2} - A_{1}^{2} \right) \right\} \, ,
\ee
$F_{1}\left(X \right)$ has extrema at 
\be
\label{eq:050410}
X = X_{\pm} \equiv 
\frac{1}{3}\left(A_{1} \pm \ \sqrt{A_{1}^{2} - 3A_{2}} \right) \, .
\ee
Because we are assuming $0 < n < 4$, we find $A_1>0$. 
Then the conditions that $\ddot a(t)$ should vanish twice 
when $0<t^2=X<t_0^2$ are given by 
\begin{align}
\label{eq:050408} 
& A_{2} > 0\, ,\quad A_{1}^{2} - 3A_{2} > 0 \, , \\
\label{eq:050409}
& F_{1}\left(X_{+} \right) < 9Ct_{1}^{2} < F_{1}\left(X_{-} \right) \, .
\end{align}
The above conditions, (\ref{eq:050408}) and (\ref{eq:050409}), means that 
the scale factor need to have the three phases for the time evolution of Universe in 
the regions $-t_{0}<t<0$ and $0<t<t_{0}$, respectively.
And its time evolution takes two deceleration and one acceleration expansions.
In this case, the Universe starts in the deceleration expansion, and take the acceleration 
phase in next, and deceleration phase in the last phase.
After that, the scale factor shrinks symmetrically, and the Universe goes to the big crunch. 

Because 
\begin{align}
\label{eq:050411}
F_{1}\left(X_{\pm} \right) =&
\frac{4-n}{108n}\left(A_{1}\left(-2A_{1}^{2} + 9A_{2} \right) \mp 
2\left(A_{1}^{2} - 3A_{2} \right) \sqrt{A_{1}^{2} - 3A_{2}} \right) \, , \\
\label{eq:050412}
A_{1}\left(-2A_{1}^{2} + 9A_{2} \right) =&
\frac{108n(8-n)}{(4-n)^{2}}t_{1}^{6}\left\{\frac{9C}{t_{1}^{4}} 
+ \frac{4n^{2}-19n+4}{n(4-n)} \right\} \, , \\
\label{eq:050413}
A_{1}^{2}-3A_{2} =&
\frac{36n}{4-n}t_{1}^{4}\left\{ \frac{-5n^{2}+20n+16}{4n(4-n)} 
 - \frac{3C}{t_{1}^{4}} \right\} \, , 
\end{align}
the conditions in (\ref{eq:050408}) give 
\begin{align}
\label{eq:050414}
\frac{C}{t_{1}^{4}} 
> \frac{n-2}{2n} \, , \\
\label{eq:050415}
\frac{-5n^{2}+20n+16}{4n(4-n)} - \frac{3C}{t_{1}^{4}} > 0 \, ,
\end{align}
and the condition (\ref{eq:050409}) gives 
\begin{align}
\label{eq:050416}
\frac{9C}{t_{1}^{4}} <&  
\frac{4-n}{108n}\left[
\frac{108n(8-n)}{(4-n)^{2}}\left\{\frac{9C}{t_{1}^{4}} 
+ \frac{4n^{2}-19n+4}{n(4-n)} \right\}
+2\left\{\frac{36n}{4-n}\left\{\frac{-5n^{2}
+20n+16}{4n(4-n)}-\frac{3C}{t_{1}^{4}} \right\} \right\}^{3/2}
\right] \, , \\
\label{eq:050417}
\frac{9C}{t_{1}^{4}} >& 
\frac{4-n}{108n}\left[
\frac{108n(8-n)}{(4-n)^{2}}\left\{\frac{9C}{t_{1}^{4}} 
+ \frac{4n^{2}-19n+4}{n(4-n)} \right\}
 -2\left\{\frac{36n}{4-n}\left\{\frac{-5n^{2}+20n+16}{4n(4-n)} 
 -\frac{3C}{t_{1}^{4}} \right\} \right\}^{3/2} \right] \, .
\end{align}
Because we are considering the case that $\frac{3}{2}<n<4$, we find 
\be
\label{nrange3}
\frac{n-2}{2n} < \frac{1}{4} \, .
\ee
Therefore Eq.~(\ref{Crange}) tells that the condition (\ref{eq:050414}) 
is always satisfied. 
We should also note that by using (\ref{Crange}), again, we obtain
\be
\label{nnrange}
\frac{-5n^{2}+20n+16}{4n(4-n)}  = \frac{5}{4} + \frac{4}{n(4-n)}
> \frac{9}{4} \geq \frac{3C}{t_{1}^{4}} \, . 
\ee
Therefore the condition (\ref{eq:050415}) is also always satisfied. 

\begin{figure}[htbp]
\centering
\includegraphics[width=7cm]{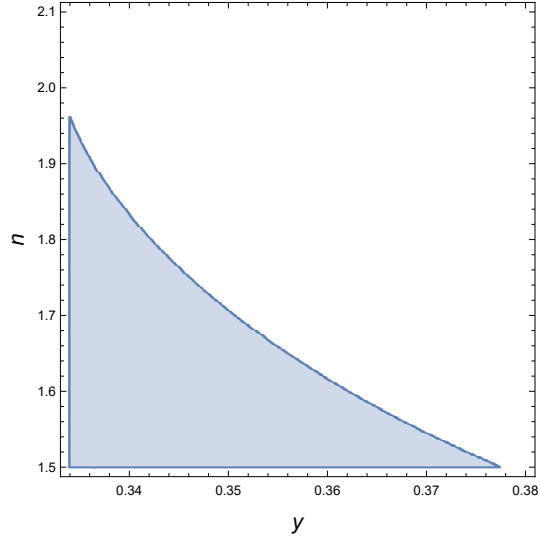}
\caption{The region satisfying the conditions (\ref{nrange}), 
(\ref{eq:050419.20.2}), and (\ref{eq:050416}) 
is shown by the colored region. }
\label{fig:fig1}
\end{figure}

By defining new functions $A_1(n,c)$ and $A_2(n,c)$ as follows, 
\begin{align}
\label{A12BBB}
& A_1(n,c) \equiv 36n \left( 4 - n \right) c+ \left( 8 - n \right) \left( 
4 n^2 - 19 n + 4 \right) \, , \nn
& A_2(n,c) \equiv \frac{1}{2} A_0(n,c)^{\frac{3}{2}} \, , \quad 
A_0(n,c) \equiv \left( 12 c - 5 \right) \left( n - 2 \right)^2 + 9 - 12 c 
\, ,
\end{align}
we rewrite the conditions (\ref{eq:050411}) and (\ref{eq:050413}) can be 
rewritten as 
\be
\label{A12B2}
A_1\left(n,c=\frac{C}{t_1^4} \right) + A_2\left( n,c=\frac{C}{t_1^4} \right) 
> 0\, , \quad 
A_1\left(n,c=\frac{C}{t_1^4} \right) - A_2\left( n,c=\frac{C}{t_1^4} \right) 
<0 \, .
\ee
We now show that the second condition in (\ref{A12B2}) or (\ref{eq:050413}) 
is always satisfied if we assume (\ref{nrange}) and (\ref{Crange}). 
We should note $A_1(n,c)$ has a local minimum with respect $n$ at $n=2$ and 
a local maximum at $n=\frac{13}{2} - 6c$. 
When $c=\frac{C}{t_1^4}$, by using (\ref{Crange}), we find 
\be
\label{1326c}
2< \frac{13}{2} - 6c < 4 \, ,
\ee
and therefore the local maximum appears in the range of (\ref{nrange}). 
We should note that 
\be
\label{A1s}
A_1(4,c)= - 32 \, , \quad - 48 < A_1(2,c) = 144c - 108 < 76 \, , \quad 
 - 47 < A_1\left(\frac{3}{2},c\right) = 135c - \frac{403}{4} < \frac{1}{2} \, ,
\ee
and 
\be
\label{An1326c}
A_1 \left(\frac{13}{2} - 6c,c \right) = \frac{297}{4} - 585c + 972 c^2 
 - 432 c^3 \, .
\ee
The r.h.s. of (\ref{An1326c}) has a local minimum $-32$ at $c=\frac{5}{12}$ 
and a local maximum at $c=\frac{13}{12} > \frac{3}{4}$. 
We should also note the r.h.s. of (\ref{An1326c}) vanishes at $c=\frac{3}{4}$. 
Therefore we find 
\be
\label{297etc}
 -32 <\frac{297}{4} - 585c + 972 c^2 - 432 c^3 < 0\, .
\ee
and therefore when $c=\frac{C}{t_1^4}<\frac{403}{540}$, $A_1(n,c)$ is always 
negative. 
Because $A_2(n,c)$ is always positive, as long as 
$c=\frac{C}{t_1^4}<\frac{403}{540}$, the second condition 
in (\ref{A12B2}) or (\ref{eq:050413}) is always satisfied. 
By using the numerical calculation, as long as  the conditions (\ref{nrange}), 
(\ref{eq:050419.20.2}), and (\ref{Crange}) are satisfied, we confirm that  the second condition 
in (\ref{A12B2}) or (\ref{eq:050413}) is always satisfied.
On the other hand, the region satisfying the condition (\ref{eq:050416}) 
is specified by the colored region in FIG.~\ref{fig:fig1}. 

\begin{figure}[htbp]
\centering
\includegraphics[width=4.8cm]{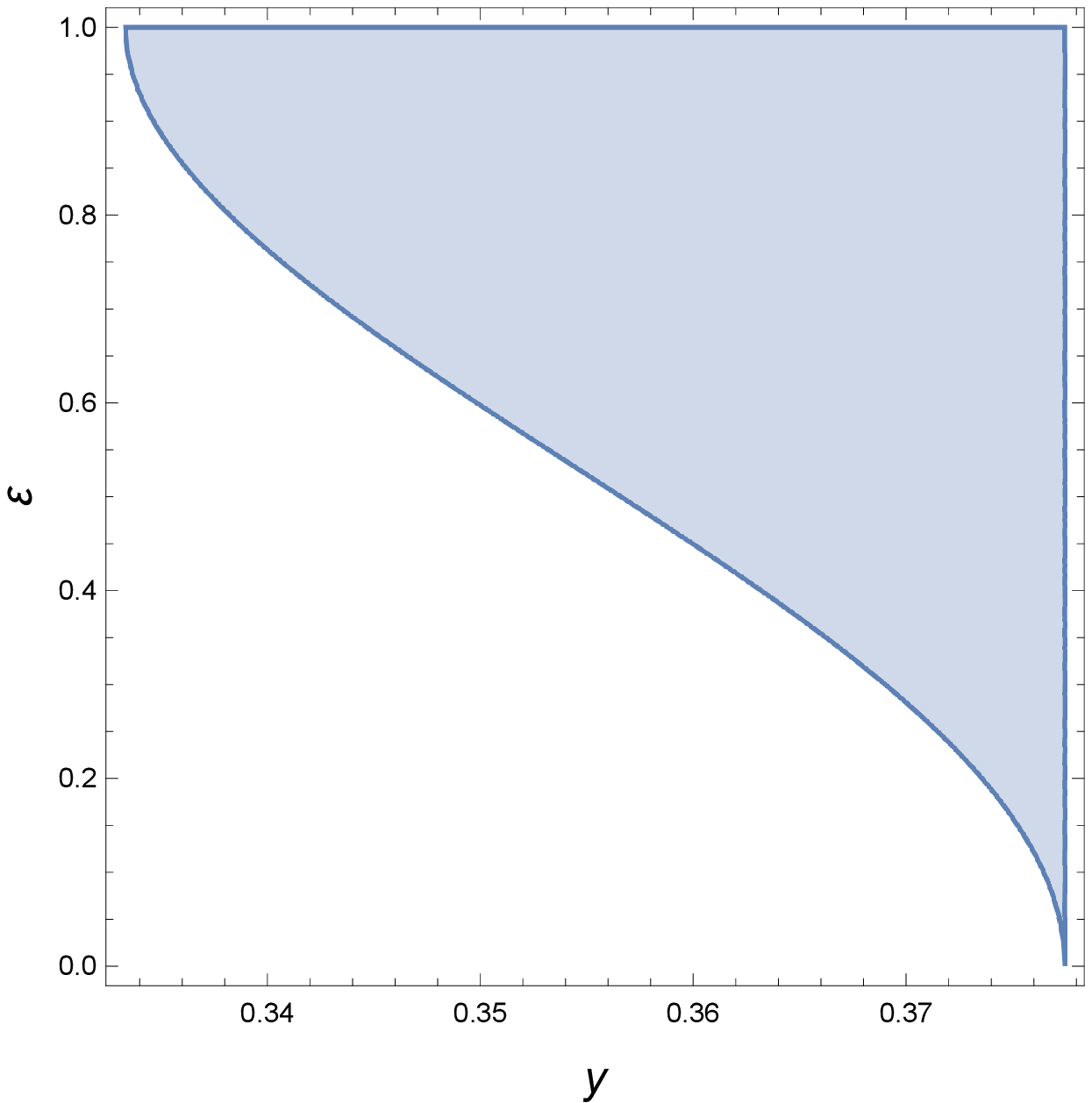}
\includegraphics[width=4.8cm]{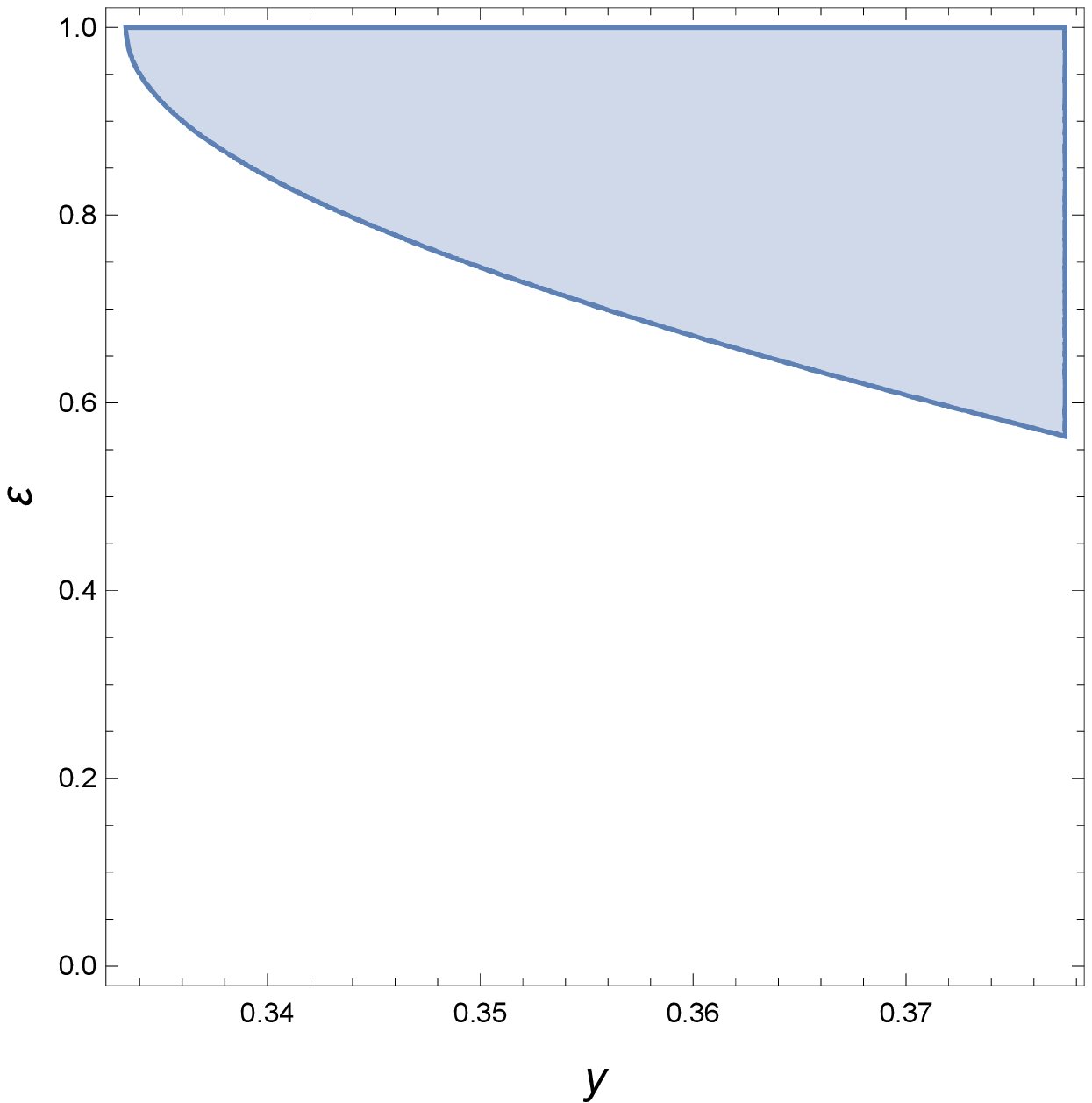}
\includegraphics[width=4.8cm]{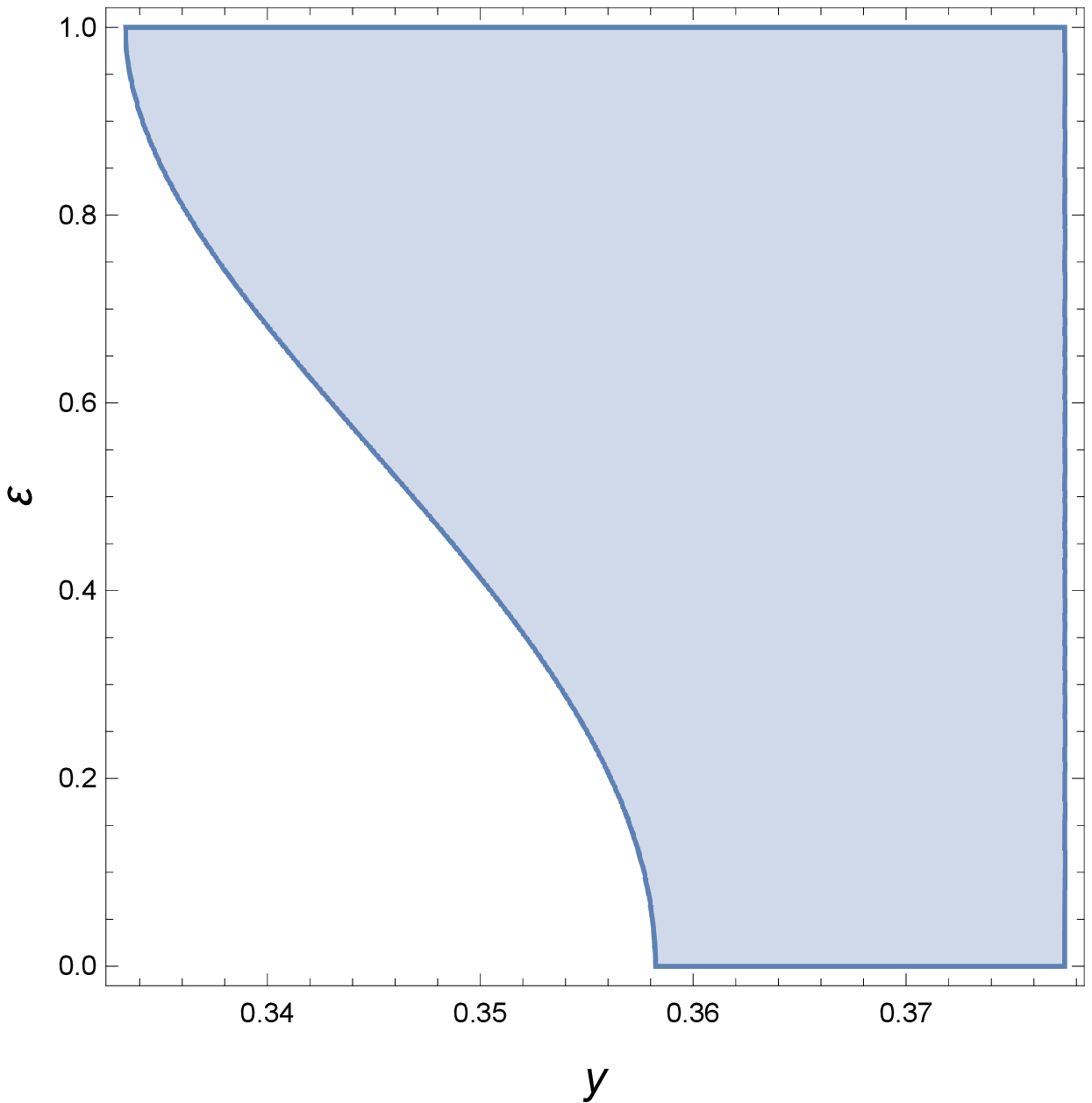}
\caption{The region satisfying the condition (\ref{ht0}) at $n=3/2$ 
is given by the colored region. 
Let $y_\mathrm{Max}$ the maximum value of $y$ which satisfies 
all the conditions. 
The left graph corresponds to $1/z=D(0,y_\mathrm{Max})$, 
the middle one to $1/z>D(0,y_\mathrm{Max})$, 
and the right graph to $1/z<D(0,y_\mathrm{Max})$. 
}
\label{fig:fig2}
\end{figure}

In case of $n=3/2$, we need to include the condition (\ref{nrange2}). 
By defining $z\equiv \frac{1}{t_{0}^{2}H_{0}^{2}}$ and 
$\epsilon\equiv \left| t_{p}/t_{0} \right|\ (0<\epsilon<1)$ with the present 
time $t_p$, the condition (\ref{nrange2}) for $h(t_{0}) > 0$ when $n=3/2$ 
can be rewritten as 
\be
\label{ht0}
D(\epsilon,y)\equiv
\frac{16}{81\Omega_{m0}}\left(\frac{3y-1}{1-\epsilon^{2}}\right)^{2}
\left(y-\frac{1-\epsilon^{4}}{6(1-\epsilon^{2})}\right)^{-2}
>\frac{1}{z} \, .
\ee
Then we find that the colored region in 
FIG.~\ref{fig:fig2} satisfies the necessary conditions. 

\begin{figure}[htbp]
\centering
\includegraphics[width=7cm]{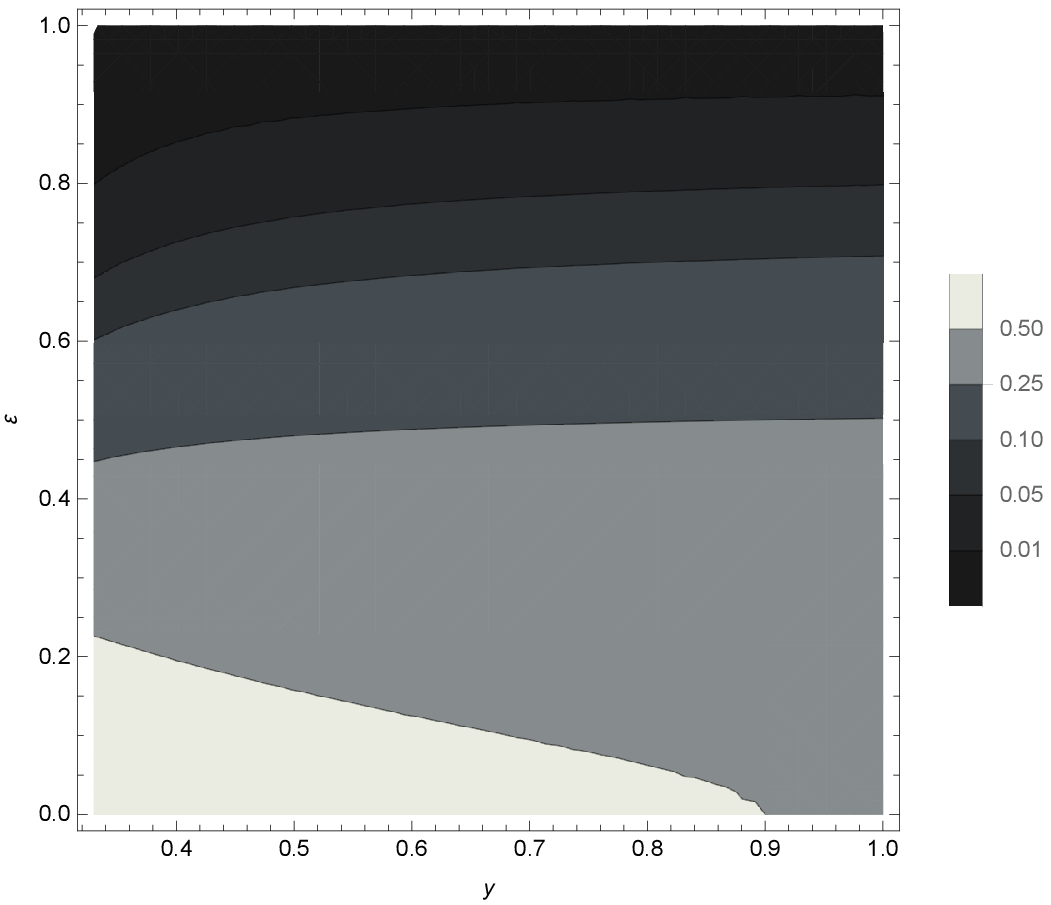}
\includegraphics[width=7cm]{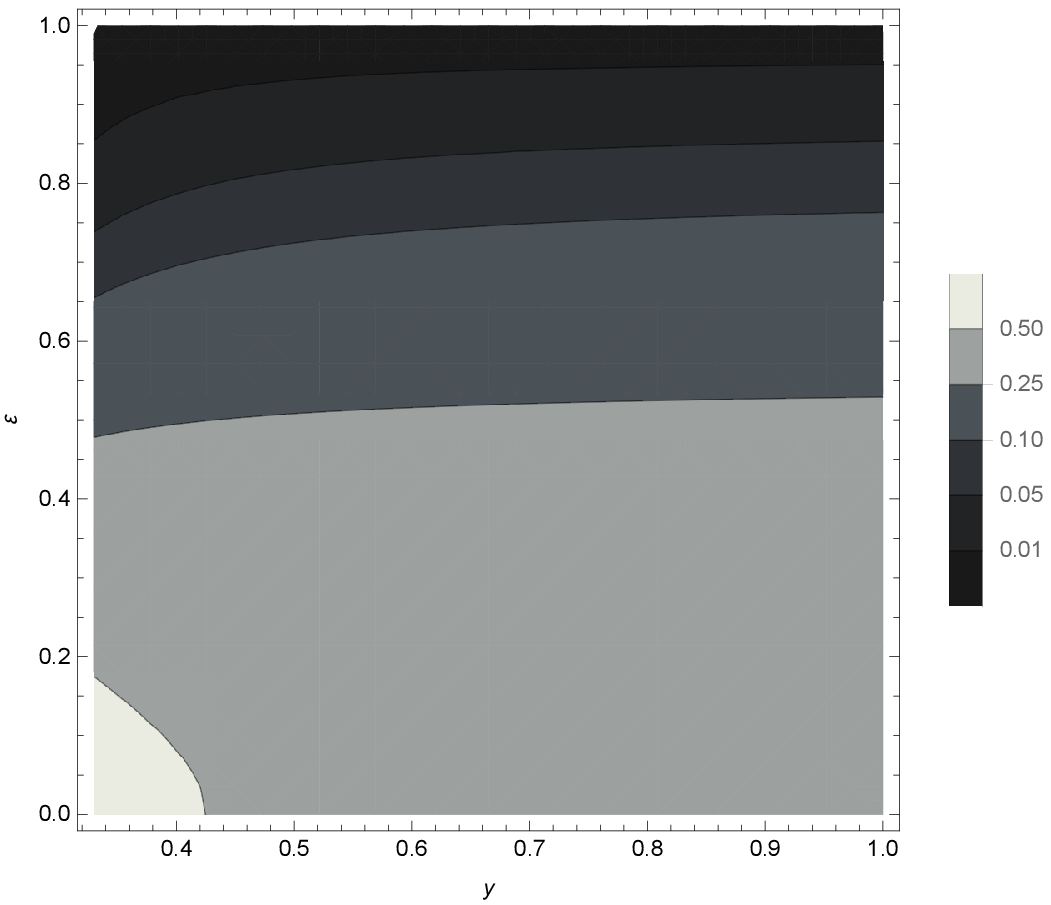}
\caption{
These graphs show the value of $\Lambda_{\mathrm{eff}}/\rho_{m0}$ 
at $n=1.5, n=2.0$ in $(y,\epsilon)$ plane.
$\Lambda_{\mathrm{eff}}/\rho_{m0}(n,y,\epsilon) \leq 
\Lambda_{\mathrm{eff}}/\rho_{m0}(1.5,0.33,0.0) \simeq 0.62$, 
the ratio of (\ref{eq:050443})  to $\rho_{m0}$ is less than 1.
The contribution of $\Lambda_{\mathrm{eff}}$ is suppressed in the order of a matter contribution.
}
\label{fig:fig3}
\end{figure}

We now estimate $\Lambda_\mathrm{eff}$ in the parameter region where the model becomes consistent. 
We now assume that the Universe is fulfilled with nonrelativistic matters 
(dust) and neglect the contribution from the relativistic matter (radiation). 
Then we find 
\be
\label{eq:050443}
\left|\frac{1}{4}\langle\tau^{\alpha}_{\ \alpha} \rangle\right|
= \frac{\rho_{m0}}{4} d\left(y, n, \epsilon \right)\left[\int^{1}_{0}
\left(x^{4}-6yx^{2}+6y-1 \right)^{3/n} dx \right]^{-1} \, , \quad 
d\left(y, n, \epsilon \right) \equiv \left\{6(1-\epsilon^{2})
\left(y-\frac{1-\epsilon^{4}}{6(1-\epsilon^{2})} \right) \right\}^{3/n} \, .
\ee
Here $x=t/t_0$ $\left(0<x<1 \right)$.
As shown in FIG.~ \ref{fig:fig3}, by adjusting the value of $\epsilon$, we can choose 
$\Omega_{\Lambda_\mathrm{eff}}$ to be less than $\Omega_\mathrm{m}$. 
However, from FIG.~ \ref{fig:fig3}, we can see that the ratio of (\ref{eq:050443})  
to $\rho_{m0}$ is less than unity in all range of $\epsilon$ and in $y$'s range.
So we can always choose the parameters for each $n$ in which no large contribution 
to $\Lambda_{\mathrm{eff}}$ happen again.
In $n=3/2$, there is the relations between $\epsilon$ and $y$.
In this case, however, we always can choose the parameters like that, too. 

\section{Summary and Conclusion \label{SecV}}

In this paper, we have introduced the scalar-tensor theory in order to realize the sequestering mechanism in the acceleratingly expanding Universe 
because the original model of the sequestering mechanism cannot realize the accelerating expansion.
The reason why the sequestering mechanism cannot realize the accelerating expansion in the original model is 
that the effective cosmological constant $\Lambda_{\mathrm{eff}}$ becomes negative and works as the negative cosmological constant in the general relativity.
Thus, we need to introduce other candidates of dark energy. 
We also have found that even though the gravitational model is modified, 
the large contribution from the quantum corrections of the matter sector
to the vacuum energy is canceled out in the same way as in the original sequestering mechanism. 
The effective cosmological constant $\Lambda_{\mathrm{eff}}$ is also described by the global average of the trace of energy-momentum tensor.

We have also estimated the value of $\Lambda_{\mathrm{eff}}$ in an example of the models.
Due to the reconstruction, 
all physical quantities can be described by the scale factor.
Thus, when we give the form of the scale factor, 
we can determine the time evolution of these quantities and evaluate these quantities. 
Because the given scale factor needs to be consistent with the cosmological history, 
and there should not exist the ghost mode, 
we have restricted the parameter regions of the scale factor.
In our model, we have found the parameter regions to make $\Lambda_{\mathrm{eff}}$, 
given by Eqs.~(\ref{lambda_eff}) and (\ref{eq:050443}), 
be less than the value of the present energy density of matters.
So, we have obtained the solutions of the scale factor with a small $\Lambda_{\mathrm{eff}}$.
In Sec.~\ref{SecIV}, we have introduced a scalar field as the source of dark energy.
And we have found the above residual term $\Lambda_{\mathrm{eff}}$ also work as dark energy.
So we may 
regard the total dark energy as the sum of $\Lambda_{\mathrm{eff}}$ 
and the energy density of scalar field.
In this paper, we have reconstructed the motion of the scalar field and set up the model 
to give the present Universe.
The scalar field takes on the part of dark energy apart from that of $\Lambda_{\mathrm{eff}}$, 
and the total energy density of dark energy takes the present observational value.
We have found that the energy density of the total dark energy is comparable to that of 
the matter sector at the present time. 

In the formulation of the present work, however, 
we need to assume the form of the scale factor, 
and then we need to determine the time evolution of the future.
This is problematic because we cannot know the future.
So, we need to solve the differential equation for the scale factor under a suitable condition.
After that, we can determine the value of $\Lambda_{\mathrm{eff}}$.

As it has been pointed out in Sec.~\ref{SecIV}, 
we have assumed the model
where the spatial volume and lifetime of the Universe are finite,
so that the four-dimensional space-time average as in 
Eq.~(\ref{lambda_average}) could be well defined.
And then, the spatial curvature does not vanish because the spatial volume is finite. 

The observation of cosmic microwave background (CMB) tells us 
that the spatial curvature should be negligibly small, 
which suggests that the radius of the Universe was large enough 
and the spatial curvature was small enough 
when the Universe became transparent to radiation. 
This may give us some constraints on the model.
The consistent inflation scenario with the sequestering mechanism
was discussed in \cite{Kaloper:2014dqa},
and one can expect that the CMB power spectrum would be obtained
in the ordinary manner after the quantum corrections are separated from the matter field
in Eq.~(\ref{separated_energy_momentum_tensor}).

Instead of the case that the lifetime of the Universe is finite, 
one can consider an alternative scenario that the Universe has a periodicity in time; 
that is, the cyclic Universe. 
By following the ekpyrotic scenario, where the hot big-bang is driven by the 
collision of the two braneworlds \cite{Khoury:2001wf}, 
it has been proposed that
the collisions occur iteratively and the Universe undergoes 
an endless sequence of cosmic epochs of each beginning with a big-bang and 
ending in a big-crunch \cite{Steinhardt:2001vw}. 
This scenario explains naturally the uniform and flat Universe with large scale structure. 
Therefore it could be interesting to embed our model into this kind of scenario.
The analysis of other models and their constrains from observational data will be treated in the future works.

\section*{Acknowledgments}

This work is supported (in part) by
MEXT KAKENHI Grant-in-Aid for Scientific Research on Innovative Areas 
``Cosmic Acceleration''  (No. 15H05890) (S.N.),
and by International Postdoctoral Exchange Fellowship Program at Central China Normal University (T.K.).

\end{document}